\documentclass[twocolumn, pra,showpacs,nofootinbib,fixfloat]{revtex4}
\usepackage{dcolumn,graphicx}
\begin{document}
\preprint{UNR Apr 2003-\today }
\title{ How to derive and compute 1,648 diagrams}
\author{Caleb C. Cannon }
\author{Andrei Derevianko}
%\email{andrei@unr.edu}
\affiliation { Department of Physics, University of Nevada, Reno,
Nevada 89557}

\date{\today}

\begin{abstract}
We present the first calculation for many-electron atoms complete
through {\em fourth} order of many-body perturbation theory.
Owing to an overwhelmingly large number of underlying diagrams,
we developed a suite of symbolic algebra tools to automate derivation and coding.
We augment all-order single-double excitation method with 1,648 omitted
fourth-order diagrams and compute amplitudes of principal
transitions in Na. The resulting {\em ab initio} relativistic electric-dipole amplitudes
are in an excellent agreement with 0.05\%-accurate experimental values. Analysis
of previously unmanageable classes of diagrams provides  a useful guide to
a design of even more accurate, yet practical many-body methods.
\end{abstract}


\maketitle

\begin{figure}[h]
\begin{center}
\includegraphics*[scale=0.5]{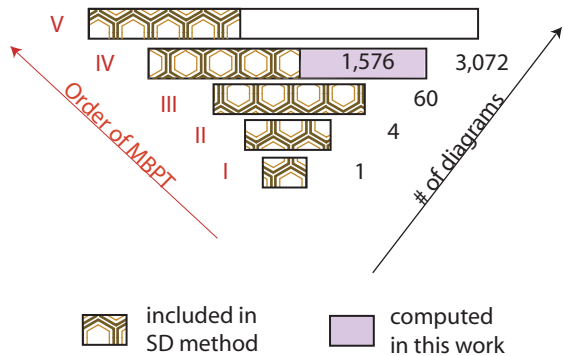}
\caption{ Number of
diagrams grows rapidly with the order of MBPT. Here we show
number of topologically distinct Brueckner-Goldstone diagrams for
transition amplitudes for univalent atoms.
We assume that calculations are carried out in $V^{N-1}$
Hartree-Fock basis to minimize the number of diagrams and we do not count
``folded''~\protect\cite{LinMor86} and normalization diagrams. All-order single-double (SD)
excitations method recovers
all diagrams through the third order, but misses roughly a half
of diagrams in the fourth order. These 1,576 missed
diagrams and  72 related normalization diagrams are explicitly computed in the present work.
  \label{Fig:CCSDvsMBPT} }
\end{center}
\end{figure}

Many-body perturbation theory (MBPT) has proven to be a powerful tool
in physics~\cite{FetWal71} and quantum chemistry~\cite{SzaOst82}.
Although MBPT provides a systematic
approach to solving many-body quantum-mechanical problem, the number
and complexity of analytical expressions and thus challenges of %computational
implementation grow rapidly with increasing order of MBPT (see Fig.\ref{Fig:CCSDvsMBPT})
Indeed, because of this complexity it has proven
to be difficult to go beyond the complete third order in calculations
for many-electron atoms (see, e.g., review \cite{Sap98}). At the same time,
studies of higher orders
are desirable for improving accuracy of {\em ab initio}
atomic-structure methods. Such an improved accuracy is
required, for example, in interpretation of
atomic parity violation~\cite{PNCreviews} and unfolding cosmological evolution of the
fine-structure constant $\alpha$~\cite{WebMurFla01}.

Here we report the first calculation
of transition amplitudes for alkali-metal atoms complete
through the fourth order of MBPT. We explicitly
computed 1,648 topologically distinct Brueckner-Goldstone diagrams.
To overcome
such an overwhelming complexity we developed a symbolic
problem-solving environment that automates highly repetitive
but error-prone
derivation and coding of many-body diagrams.
Our work illuminates a crucial role of symbolic tools
in problems unsurmountable in the traditional ``pencil-and-paper'' approach.
Indeed, third-order calculation~\cite{BluGuoJoh87,JohLiuSap96}
was a major research project most likely to have required a year to
accomplish. As one progresses from the third to the present fourth order (see Fig.~\ref{Fig:CCSDvsMBPT}) there is a 50-fold
increase in the number of diagrams. Simple scaling shows
that present calculations require {\em half a century}  to complete.
With our tools derivation and coding take just a few minutes.
Similar symbolic tools were developed by \citet{PerXiaFlu01},
however their package is so far limited to well-studied\cite{BluJohSap90a,JohLiuSap96}
third order of MBPT. In contrast,
we explore a wide range of new, previously unmanageable, classes of diagrams.

As an example of application of our symbolic technology,
we compute   electric-dipole amplitudes
of the principal $3p_{3/2}-3s_{1/2}$ and $3p_{1/2}-3s_{1/2}$  transitions in Na.
We augment all-order single-double excitations method~\cite{SafDerJoh98}
with  1,648 diagrams so
that the formalism is complete through the fourth order ( see Fig.\ref{Fig:CCSDvsMBPT} ).
The results are in  excellent agreement with 0.05\%-accurate
experimental values~\cite{JohJulLet96}.
Thus our computational method not only enables exploration of a wide range of previously
unmanageable classes of diagrams but also defines new level of accuracy in {\em ab initio} relativistic
atomic many-body calculations. Atomic units $|e|=\hbar=m_e=4\pi\varepsilon_0\equiv 1$ are used
throughout this paper.

{\it Method---}
A practitioner
of atomic MBPT typically follows these steps: (i) derivation of diagrams, (ii) angular reduction, and
(iii) coding and numerical evaluation. Below we highlight components of our
problem-solving environment designed to assist
a theorist in these technically-involved tasks.
First we briefly reiterate MBPT formalism~\cite{DerEmm02}
for atoms with a single valence electron outside a closed-shell
core. For these systems a convenient point of departure
is a single-particle basis generated in frozen-core ($V^{N-1}$) Dirac-Hartree-Fock (DHF)
approximation~\cite{Kel69}. In this approximation the number of  MBPT diagrams
is substantially reduced~\cite{LinMor86,BluGuoJoh87}.
The lowest-order valence wavefunction
$|\Psi_v^{(0)}\rangle $ is simply a Slater determinant constructed
from core orbitals and proper valence state $v$.
The perturbation
expansion is built in powers of  residual interaction $V_I$ defined as
a difference between the full Coulomb interaction between
the electrons and the DHF potential.
The $n^\mathrm{th}$-order correction to the valence wavefunction may be expressed as
\begin{equation}
|\Psi_v^{(n)}\rangle = -R_{v}\left\{  Q\,V_I\,|\Psi_v^{(n-1)}\rangle\right\}
_{\mathrm{linked}} \, ,
\label{Eq:Psin}
\end{equation}
where $R_v$ is a resolvent operator modified to include so-called ``folded''
diagrams~\cite{DerEmm02}, projection operator $Q=1-|\Psi_v^{(0)}\rangle \langle \Psi_v^{(0)}| $,
and only linked diagrams~\cite{LinMor86} are to be kept.
From this recursion relation we may generate corrections to wave functions at any
given order of perturbation theory. With such calculated
corrections to wavefunctions of two valence
states $w$ and $v$, $n^\mathrm{th}$-order contributions to matrix
elements of an operator $\hat{Z}$ are
\begin{equation}
Z^{(n)}_{wv} = \sum_{k=0}^{n-1}
\langle \Psi_w^{(n-k-1)} |  Z  |\Psi_v^{(k)}  \rangle_\mathrm{val,\,conn}
+ Z^{(n)}_{wv, \, {\rm norm}}\label{Eq:Zn}     \, .
\end{equation}
Here $Z^{(n)}_{wv, \, {\rm norm}}$ is a normalization correction arising due to an intermediate
normalization scheme employed in derivation of Eq.~(\ref{Eq:Psin}). Subscript
``$\mathrm{val,\,conn}$'' indicates that only connected diagrams involving excitations
from valence orbitals are included in the expansion.

Equations~(\ref{Eq:Psin}) and (\ref{Eq:Zn}) completely define a set
of many-body diagrams at any given order of MBPT. In practice the derivations
are carried out in the second quantization
and the Wick's theorem~\cite{LinMor86} is used to simplify products of
creation and annihilation operators.
Although the application of the Wick's theorem is straightforward, as order of
MBPT increases, the sheer length of expressions and number of operations
becomes quickly unmanageable. We developed a symbolic-algebra
package written in {\em Mathematica}~\cite{Wol99} to carry out this task.
The employed algorithm relies on decision trees and pattern matching, i.e.,
programming elements typical to artificial intelligence applications.
With the developed package we fully derived fourth-order corrections to matrix elements
of univalent systems~\cite{DerEmm02}.

This is one of the fourth-order terms  from Ref.~\cite{DerEmm02}
\begin{equation}
\sum_{abc} \sum_{mnr} \frac{{z_{bv}}{\tilde{g}}_{canr}{g_{nrcm}}{\tilde{g}}_{mwab}}
     {({{\varepsilon }_w}-{{\varepsilon }_{b}})\,
       ({{\varepsilon }_{mw}}-{{\varepsilon }_{ab}})\,
       ({{\varepsilon }_{nrw}}-{{\varepsilon }_{abc}})}\, \, .
       \label{Eq:sample}
\end{equation}
There are 524 such contributions
in the fourth order~\cite{nmbDiag:comment}.
Here abbreviation
$\varepsilon_{xy\ldots z}$  stands for
$\varepsilon_{x} + \varepsilon_{y} + \cdots \varepsilon_{z}$, with $\varepsilon_{x}$
being single-particle DHF energies.
Further, $g_{ijlk}$ are matrix elements of electron-electron interaction in the basis
of DHF orbitals.
The quantities $\tilde{g}_{ijlk}$
are antisymmetric combinations $\tilde{g}_{ijlk}={g}_{ijlk}- {g}_{ijkl}$.
The summation is over single-particle DHF states.
Core orbitals are enumerated by letters $a,b,c$ and complementary
excited states are labelled by $m,n,r$.
Finally matrix elements of the operator $\hat{Z}$ in the DHF basis are denoted as $z_{ij}$.

The summations over magnetic quantum
numbers are usually carried out analytically.  This ``angular reduction''
is the next major technically-involved step. We also automate this task.
The details are provided in Ref.~\cite{DerAng02}.
Briefly, the angular reduction is based on application of the Wigner-Eckart (WE) theorem~\cite{Edm85}
to matrix elements $z_{ij}$ and $g_{ijkl}$. The WE theorem allows one to ``peel off''
dependence of the matrix elements on magnetic quantum numbers in the form of 3j-symbols
and reduced matrix elements.
In the particular case of fourth-order terms, such as Eq.~(\ref{Eq:sample}), application of the WE theorem
results in a product of seven 3j-symbols.
To automate simplification of the products of 3j-symbols we  employed a symbolic  program
{ Kentaro} developed by \citet{Tak92}.

The result of angular reduction of our sample term~(\ref{Eq:sample}) is
\begin{eqnarray*}
\lefteqn{ \sum_{abcmnr}\sum_{L}
     \frac{\delta_{j_a j_m}\,\delta_{j_b j_w}\,
        {\left( -1 \right) }^{j_a + j_c + j_n + j_r}\,}
        {(2L+1)\,{\sqrt{(2j_a+1) (2j_w+1)}}}}\\
&&   \frac{\langle b||z||v\rangle \,Z_{L}(canr) X_{L}(nrcm)\,
        Z_{0}(mwab)\,}{
        ({{\varepsilon }_w}-{{\varepsilon }_{b}})\,
        ({{\varepsilon }_{mw}}-{{\varepsilon }_{ab}})\,
        ({{\varepsilon }_{nrw}}-{{\varepsilon }_{abc}})} \, .
\end{eqnarray*}
Here the reduced quantities $\langle i|| z || j \rangle$,
$X_{L}(ijkl)$, and $Z_{L}(ijkl)$
depend only on total angular momenta and principal quantum numbers of single-particle orbitals.

As a result of angular reduction we generate analytical expressions suitable for
coding. We also automated the tedious coding process by developing custom parsers
based on Perl and Mathematica. These parsers translate analytical expressions into
Fortran90 code.
The resulting code is very large - it is about 20,000 lines long
and were it be programmed manually,  it would have required several years to develop.
For numerical evaluation we employed a  B-spline library\cite{JohBluSap88}.
All the fourth-order results were computed with a sufficiently large basis of
25 out of 30 lowest-energy ($E>mc^2$) spline functions for each partial wave through $h_{11/2}$.

At this point we have demonstrated  feasibility of working with thousands of diagrams
in atomic MBPT. Now we apply our computational
technique to high-accuracy calculation of transition amplitudes in Na.

{\em Fourth-order diagrams complementary to single-double excitations method.}
One of the mainstays of practical applications of MBPT is an assumption
of convergence of series in powers of the perturbing
interaction. Sometimes the convergence is poor and then one
sums certain classes of diagrams  to ``all orders''
using iterative techniques. In fact, the most accurate
many-body calculations of parity violation in Cs by \citet{DzuFlaSus89} and \citet{BluJohSap90}
are of this kind. These techniques, although summing certain classes of MBPT diagrams
to all orders,
still  do not account for  an infinite number of residual diagrams (see Fig.~\ref{Fig:CCSDvsMBPT}).
In Ref.~\cite{DerEmm02} we proposed to augment a given all-order technique with some of the omitted
diagrams so that the formalism is complete through a certain order of MBPT.
As in that work, here we consider an improvement of all-order single-double (SD) excitation
method
employed in Ref.~\cite{BluJohSap90}. Here a certain level $n$ of
excitations from lowest-order wavefunction  refers to an all-order grouping of
contributions in which $n$ core and valence electrons are promoted to excited single-particle
orbitals. The SD method is a simplified version of
the coupled-cluster expansion truncated at single and double excitations.

The next step in improving the SD method would be an inclusion of triple
excitations.
However, considering present state of available computational power,
 the complete incorporation of triples
seems as yet impractical for heavy atoms. Here we investigate an alternative illustrated in Fig.~\ref{Fig:CCSDvsMBPT} :
we compute the dominant contribution of triples in
a direct fourth-order MBPT for transition amplitudes. We also
account for contribution of disconnected quadruple excitations
in the fourth order.
In Ref.~\cite{DerEmm02}, we separated these complementary diagrams into three major categories by noting
that triples and disconnected quadruples enter the fourth order matrix element $Z^{(4)}_{wv}$  via
(i) an {\em indirect} effect of triples and disconnected quadruples on single and double excitations in the
third-order wavefunction --- we denote this class as $Z_{0\times 3}$;
(ii) {\em direct} contribution  to matrix elements
labelled  as $Z_{1\times 2}$;
(iii) correction to normalization denoted as $Z_\mathrm{norm}$.
Further these categories were broken into subclasses based on the nature of triples, so that
\begin{eqnarray}
\lefteqn{ \left( Z^{(4)}_{wv} \right)_\mathrm{non-SD} =
Z_{1 \times 2}(T_v) + Z_{1 \times 2}(T_c) + } \nonumber \\
& &Z_{0 \times 3}(S_v[T_v]) + Z_{0 \times 3}(D_v[T_v]) +\\
& & Z_{0 \times 3}(S_c[T_c]) + Z_{0 \times 3}(D_v[T_c]) + \nonumber \\
& &Z_{1\times 2}\left( D_{\mathrm nl} \right) +
Z_{0\times 3}\left( D_{\mathrm nl} \right) + Z_\mathrm{norm}(T_v)  \nonumber \, .
\end{eqnarray}
Here we distinguished between valence ($T_v$) and core ($T_c$) triples and
introduced a similar notation for singles ($S$) and doubles ($D$).
Notation like $S_v[T_c]$ stands for effect of second-order core triples ($T_c$) on
third-order valence singles $S_v$. Diagrams $D_\mathrm{nl}$ are
contributions of disconnected quadruples (non-linear contributions from double excitations).
The reader is referred to Ref.~\cite{DerEmm02} for further details and discussion.
\begin{figure}[h]
\begin{center}
\includegraphics*[scale=0.5]{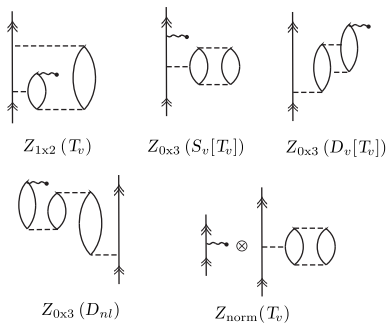}
\caption{ \small \label{FigZ4triples}
Representative fourth-order diagrams involving  triple and disconnected
quadruple excitations. }
\end{center}
\end{figure}

{\em Transition amplitudes in Na.}
Using our problem-solving environment we derived the 1,648 complementary diagrams~\cite{DerEmm02},
carried out angular reduction~\cite{DerAng02}, and generated Fortran 90 code suitable
for any univalent system.
As an example we evaluate
reduced electric-dipole matrix elements
of $3s_{1/2}-3p_{1/2,3/2}$ transitions in Na (eleven electrons)\cite{comm:choice}.
Our numerical results are presented in Table~\ref{table:D4break}.
Analyzing this table we see that leading contributions come
from valence triples $T_v$. Similar conclusion can be drawn
from our preliminary calculations for heavier Cs atom.
Dominance of valence triples ($T_v$) over core triples ($T_c$) may be explained
by smaller energy denominators for $T_v$ terms. Representative diagrams for these
relatively large contributions are shown in Fig.~\ref{FigZ4triples}.
Based on this observation we propose to fully incorporate valence triples
into a hierarchy of coupled-cluster equations and add a perturbative contributions of
core triples. Such an all-order scheme would be a more accurate and yet practical
extension of the present calculations.

Another point we would like to discuss is a sensitivity of our results to higher-order
corrections.
In Table~\ref{table:D4break}, all large contributions add up coherently,
possibly indicating a good convergence pattern of MBPT.
However, we found large, factor of 100, cancellations of terms inside the
$Z_{0 \times 3}(S_v[T_v])$ class. In principle
higher-order  MBPT corrections may offset a balance between cancelling terms
and an all-order treatment is desired. Fortunately, the $Z_{0 \times 3}(S_v[T_v])$
class of diagrams (combined with parts of $Z_{1 \times 2}(T_v)$)
have been taken into account in all-order SDpT (SD + partial triples) method~\cite{BluJohSap90,SafJohDer99}.
We correct
our results for the difference between all-order~\cite{comm:Mar}
and our fourth-order values for these diagrams (last row of Table~\ref{table:D4break}).
These all-order corrections modify our final values of complementary diagrams by  15\%.

\begin{center}
\begin{table}[h]
\begin{tabular}{lcdd}
\hline\hline
\multicolumn{1}{c}{Class } &
\multicolumn{1}{c}{Number of} &
\multicolumn{1}{r}{$3p_{1/2}-3s_{1/2}$} & \multicolumn{1}{r}{$3p_{3/2}-3s_{1/2}$}   \\
 & \multicolumn{1}{c}{diagrams} & \\
\hline
\multicolumn{4}{c}{Connected triples} \\
$Z_{0 \times 3}(S_v[T_v])$  & 72       & -0.8[-3] &  -1.1[-3]     \\
%$Z_{0 \times 3}(D_v[T_v^p])$& 216      & -1.0[-3] &   -1.5[-3]     \\
%$Z_{0 \times 3}(D_v[T_v^h])$& 216      & -1.1[-3] &   -1.6[-3]     \\
$Z_{0 \times 3}(D_v[T_v])$  & 432      & -2.2[-3] &   -3.0[-3]     \\
%$Z_{1 \times 2}(T^h_v)$     &252       &  8.4[-3] &   11.9[-3]     \\
%$Z_{1 \times 2}(T^p_v)$     & 252      & -9.1[-3] &  -12.9[-3]     \\
$Z_{1 \times 2}(T_v)$     & 504      & -0.7[-3] &   -1.0[-3]     \\[0.5ex]
$Z_\mathrm{norm}(T_v)$       & 72       & -0.7[-3] &   -1.2[-3]     \\
$Z_{0 \times 3}(D_v[T_c])$  & 144      & -0.01[-3]&   -0.01[-3]     \\
$Z_{0 \times 3}(S_c[T_c])$  & 72       &  0.06[-3]&    0.09[-3]    \\
$Z_{1 \times 2}(T_c)$       & 216      &  0.03[-3]&   0.04[-3]      \\
Total triples               & 1512     & -4.3[-3] &   -6.3[-3] \\[1ex]
\multicolumn{4}{c}{Disconnected quadruples} \\
$Z_{0 \times 3}(D_{nl})$    &  68       & 1.1[-3] &  1.6[-3]  \\
$Z_{1 \times 2}(D_{nl})$    &  68       & 0.2[-3] &  0.3[-3] \\
Total quads                 & 136       & 1.4[-3] &  2.0[-3] \\
\hline
Total                       & 1648      & -2.6[-3] & -4.3[-3] \\
+ $\delta$(SDpT)                  &           & -3.3[-3] & -4.9[-3]  \\
\hline\hline
\end{tabular}
\caption{Fourth-order complementary contributions to
reduced electric-dipole matrix elements $\langle 3p_j||D||3s_{1/2} \rangle$ in Na.
Last row marked  ``+ $\delta$(SDpT)''  is the total
value corrected using all-order SDpT values as discussed in the text.
Notation $x[y]$ stands for $x \times 10^y$.
\label{table:D4break}
}
\end{table}
\end{center}

In Table~\ref{table:comparison} we add our complementary
diagrams to SD matrix elements~\cite{SafDerJoh98} and compare with experimental
values. Several high-accuracy experiments
have been carried out for Na, resolving an apparent disagreement
between an earlier measurement and calculated lifetimes~\citep[see review][and references therein]{VolSch96}.
In Table~\ref{table:comparison} we compare with the results of the two most accurate
experiments\cite{JohJulLet96,VolMajLie96}. The SD
method~\cite{SafDerJoh98} overestimates these experimental
values by 2.5 $\sigma$ and 2.8 $\sigma$ respectively ($\sigma$ is experimental
uncertainty).  With our fourth-order corrections taken into consideration the comparison
significantly improves.
The resulting {\em ab initio} matrix elements for both $3p_{1/2}-3s_{1/2}$ and $3p_{3/2}-3s_{1/2}$
transitions are
in an excellent agreement with 0.05\%-accurate values from Ref.~\cite{JohJulLet96}
and differ by 1.2$\sigma$ from less-accurate results of Ref.~\cite{VolMajLie96}.
Considering this agreement it would be desirable to have experimental data accurate to 0.01\%.

\begin{center}
\begin{table}[h]
\begin{tabular}{ldd}
\hline\hline
\multicolumn{1}{c}{ } &
\multicolumn{1}{c}{$3p_{1/2}-3s_{1/2}$}   &
\multicolumn{1}{c}{$3p_{3/2}-3s_{1/2}$}   \\
\hline
Singles-doubles~\protect\cite{comm:SDvals}
                           & 3.5307  & 4.9930\\
$ \left( Z^{(4)} \right)_\mathrm{non-SD}$
                           &-0.0033  &-0.0049  \\
Total                      & 3.5274  & 4.9881\\
\multicolumn{3}{c}{Experiment}\\
\protect\citet{JohJulLet96}            &
                             3.5267(17) & 4.9875(24) \\
\protect\citet{VolMajLie96} &3.5246(23) & 4.9839(34) \\
\hline\hline
\end{tabular}
\caption{Comparison of the calculated reduced electric-dipole matrix element
$\langle 3p_j||D||3s_{1/2} \rangle$  of
principal transitions in Na with experimental data.
\label{table:comparison}}
\end{table}
\end{center}

We demonstrated that symbolic tools can replace multi-year detailed development
efforts in atomic MBPT with an interactive work at a conceptual level,
thus enabling an exploration of ever more sophisticated techniques.
As an example, we presented the first calculations for many-electron atoms complete
through fourth order, a task otherwise requiring half a century to complete.
Although even at this level the computed transition amplitudes for Na
indicate a record-setting {\em ab initio}
accuracy of a few 0.01\%,
the calculations allowed us to gain
insights into relative importance of various contributions and to propose
even more accurate yet practical many-body method.
With an all-order generalization~\cite{DerEmm02} of the derived diagrams
we plan to address a long-standing problem \cite{DzuFlaSus89,BluJohSap90} of improving theoretical
accuracy of interpretation of parity violation in Cs atom~\cite{WooBenCho97}.

We would like to thank W.R. Johnson, V.A. Dzuba, W.F.  Perger, and K. Takada  for
useful discussions and M.S. Safronova for providing
detailed breakdown of SDpT and SD results.
This work was supported in part by the National Science Foundation.

\bibliography{add2400,pnc,general,mypub,exact,lifetimes}

\begin{thebibliography}{29}
\bibitem[{nmb()}]{nmbDiag:comment}
\bibinfo{note}{This number does not include folded diagrams or normalization
  correction. In contrast to Ref.~\cite{DerEmm02} we have also counted
  so-called Hermitian conjugated terms as a separate contribution. When
  anisymmetrized Coulomb matrix elements $\tilde{g}_{ijkl}$ are expanded,
  term~(\ref{Eq:sample}) is represented by four topologically unique
  Brueckner-Goldstone diagrams.}
\bibitem[{com({\natexlab{a}})}]{comm:choice}
\bibinfo{note}{Several high-accuracy experimental data sets are available for
  these transitions and, unlike hyperfine-structure constants, an
  interpretation is free from QED and nuclear structure uncertainties}.
\bibitem[{com({\natexlab{c}})}]{comm:SDvals}
\bibinfo{note}{We recomputed SD values to more significant digits compared to
  Ref.~\protect\cite{SafDerJoh98}}.
\end{thebibliography}

\end{document}